\documentclass[aps,prl,reprint,groupedaddress]{revtex4-1}

\usepackage{verbatim}
\usepackage{latexsym,amssymb,float}
\usepackage{setspace}
\usepackage{graphicx}
\usepackage{epsfig}
\usepackage{amsmath}
\usepackage{bm}

\begin{document}

\title{Scanning Tunneling Potentiometry, Charge Transport and Landauer's Resistivity Dipole from the Quantum to the Classical Transport Regime}
\author{Dirk K. Morr}
\affiliation{University of Illinois at Chicago, Chicago, IL 60607, USA}

\date{\today}

\begin{abstract}
Using the non-equilibrium Keldysh formalism, we investigate the spatial relation between the electro-chemical potential measured in scanning
tunneling spectroscopy, and local current patterns over the entire range from the quantum to the classical transport regime. These quantities show similar spatial patterns near the quantum limit, but are related by Ohm's law only in the classical regime. We demonstrate that defects induce a Landauer residual resistivity dipole in the electro-chemical potential with the concomitant spatial current pattern representing the field lines of the dipole.

\end{abstract}

\pacs{}

\maketitle

\begin{figure}
 \includegraphics[width=8cm]{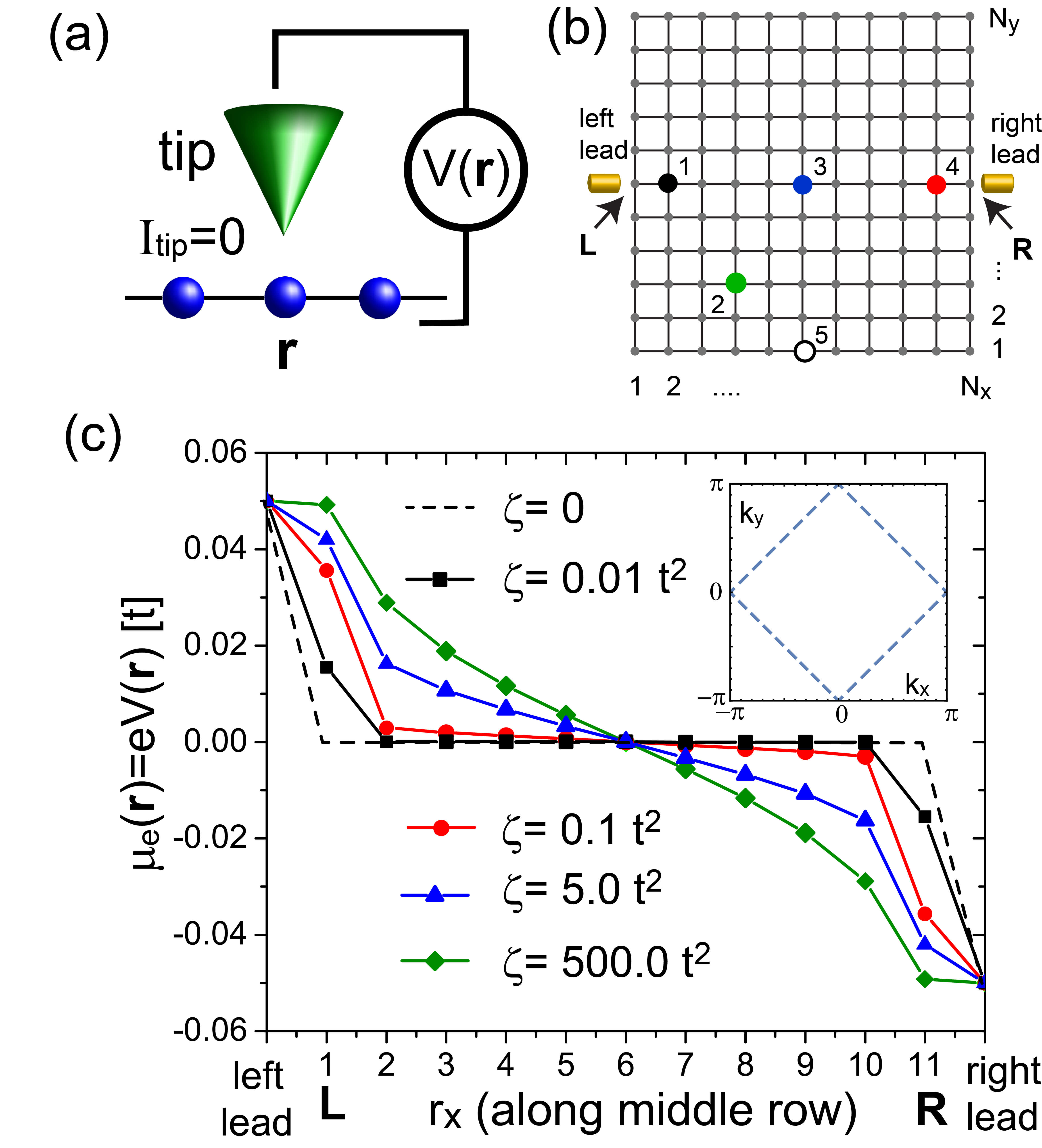}%
 \caption{(a) Schematic representation of STP: when the STP tip is above at site ${\bf r}$ of the network, its potential $V({\bf r})$ is adjusted such that there is a zero net current flowing between the tip and the network. (b) Network of electronic sites that are connected by electronic hopping (solid black lines) and coupled to two narrow leads. (c) $\mu({\bf r})$ along the middle row of the network in (b) for different values of $\zeta$. Inset: Fermi surface of the network.}
 \label{fig:mu_cuts}
 \end{figure}

Visualizing charge transport at the nanoscale is not only of great fundamental interest to understand and explore the crossover from quantum to diffusive transport, but also important for the continued miniaturization of electronic circuits. While spatial imaging of charge currents at the meso-scale has been achieved using scanning probe microscopy \cite{Cro00a,Cro00b,Eri96,Top00,Top01,Aid07,Jura07,Hac06,Mar07,Hac10}, scanning tunneling potentiometry \cite{Mur86,Bri96,Chu90,Wang10} (STP) [see Fig.~\ref{fig:mu_cuts}(a)] has been employed to gain insight into nature of charge transport at much smaller length scales down to the nanometer scale \cite{Hom09,Ji12,Wang13,Wil15}. This has led to the observation of Landauer's residual resistivity dipoles \cite{Lan57,Lan75} near step edges \cite{Hom09,Wang13,Wil15}. The question, however, arises of whether one can gain direct insight into the spatial form of the current density -- or more generally the spatial current patterns --  from the electrochemical potential measured via STP. While in the limit of classical, diffusive transport, the relation between these two quantities is established by Ohm's law, most materials of interest possess sufficiently long mean free paths such that they lie either in the crossover region between classical and quantum transport, or even close to the quantum limit. In this regime, the relation between the local electro-chemical potential and the current density is unknown and identifying it is therefore crucial for visualizing the spatial flow of currents at the nanometer scale through STP.

In this article, we provide this missing link by identifying the relation between the spatial form of the electrochemical potential, $\mu_e({\bf r})$ as determined via STP and the spatial current pattern, $I_{\bf r,r^\prime}$, over the entire range from the quantum to the classical transport regime. Using the Keldysh Green's function formalism \cite{Kel65,Ram86,Car71}, we demonstrate that near the quantum limit, the spatial form of $\mu_e({\bf r})$ is similar to that of $I_{\bf r,r^\prime}$, such that the electro-chemical potential can be employed to spatially image the current pattern. On the other hand, we show that Ohm's law can only be used in the classical limit to directly deduce the local current density $I_{\bf r,r^\prime}$ from the spatial form of $\mu_e({\bf r})$. Moreover, we demonstrate that the evolution of the spatial form of the potential between the quantum and classical limit is reflected in changes of an effective Fermi distribution function. We show that defects induce a Landauer's residual resistivity dipole in $\mu_e({\bf r})$  and that the concomitant spatial form of $I_{\bf r,r^\prime}$ is that of field lines associated with the presence of a dipole. Finally, we demonstrate that $\mu_e({\bf r})$ changes sharply at interfaces or step edges accompanied by large scale spatial oscillations.   These results identify the relation between the electrochemical potential and the local flow of charges over the entire range from quantum to classical transport.

To investigate the form of local potential $\mu_e({\bf r})$, its relation to the spatial current pattern, and its evolution from the quantum to the classical limit, we consider a network of electronic sites that are connected by hopping elements as shown in Fig.~\ref{fig:mu_cuts}(b) \cite{Cre03,Tod99,Can12,Can13,Bev14,Morr16}.  These sites can represent atoms, molecules or quantum dots; for the present purpose we assume that they possess only a single electronic level. The network is coupled to two leads, and described by the Hamiltonian $H=H_0 + H_{def} + H_{ph} + H_{c} + H_{tun} + H_{tip} + H_{lead}$, where
\begin{align}
H_0 &= \sum_{{\bf r, r^\prime},\sigma} \left( -t - \mu \delta_{\bf r, r^\prime} \right) c^\dagger_{{\bf r} \sigma} c_{{\bf r^\prime} \sigma} \nonumber \\
H_{def} & = \sum_{{\bf R},\sigma} U_0 c^\dagger_{{\bf R},\sigma} c_{{\bf R},\sigma} \nonumber \\
H_{ph} & =  g \sum_{{\bf r},\sigma} c^\dagger_{{\bf r} \sigma} c_{{\bf r}\sigma} \left( a^\dagger_{\bf r} + a_{\bf r} \right) + \omega_0 \sum_{{\bf r},\sigma}  a^\dagger_{\bf r} a_{\bf r} \nonumber \\
H_c &= -t_c \sum_{j,\sigma} \left(  c^\dagger_{ {\bf R}_j,\sigma} d_{ {\bf R}_j,\sigma} + c^\dagger_{ {\bf L}_j,\sigma} d_{ {\bf L}_j,\sigma} + H.c. \right) \nonumber \\
H_{tun} &= -t_{tip} \sum_\sigma c^\dagger_{{\bf r} \sigma} f_{\sigma} + f_{\sigma}^\dagger c_{{\bf r} \sigma}  \ .
\label{eq:H}
\end{align}
Here, $c^\dagger_{{\bf r} \sigma} (c_{{\bf r^\prime} \sigma})$ creates (annihilates) an electron with spin $\sigma$ at site ${\bf r}$ in the network, $ -t$ is the electronic hopping between nearest-neighbor sites, and $\mu$ is the chemical potential. $H_{def}$ describes the electronic scattering off non-magnetic defects located at sites ${\bf R}$, and $H_{ph}$ represents the interaction of the electrons with local Einstein phonon modes of energy $\omega_0$. $H_c$ describes the coupling of the network to the left and right leads, and $H_{tun}$ represents the tunneling of an electron from the tip to a site ${\bf r}$ in the network. Finally, $H_{tip}$ and  $H_{lead}$ describe the electronic structure of the tip and the leads, respectively. Below, we assume the wide-band limit for both with a constant density of states $N_0=1/t$ and set $\mu=0$ yielding the Fermi surface shown in the inset of Fig.~\ref{fig:mu_cuts}(c). Finally, we had previously shown \cite{Morr16} that by increasing $g$, one can tune the network's transport properties from the quantum to the classical limit. To this end, we employ the high-temperature approximation $k_B T \gg \hbar \omega_0$ \cite{Bih05,Morr16} where the strength of the electron-phonon interaction is characterized by a single parameter, $\zeta=2g^2 k_B T/(\hbar \omega_0)$ with $\zeta=0$ and $\zeta \rightarrow \infty$ corresponding to the quantum and classical transport limits, respectively.
\begin{figure}
\includegraphics[width=8cm]{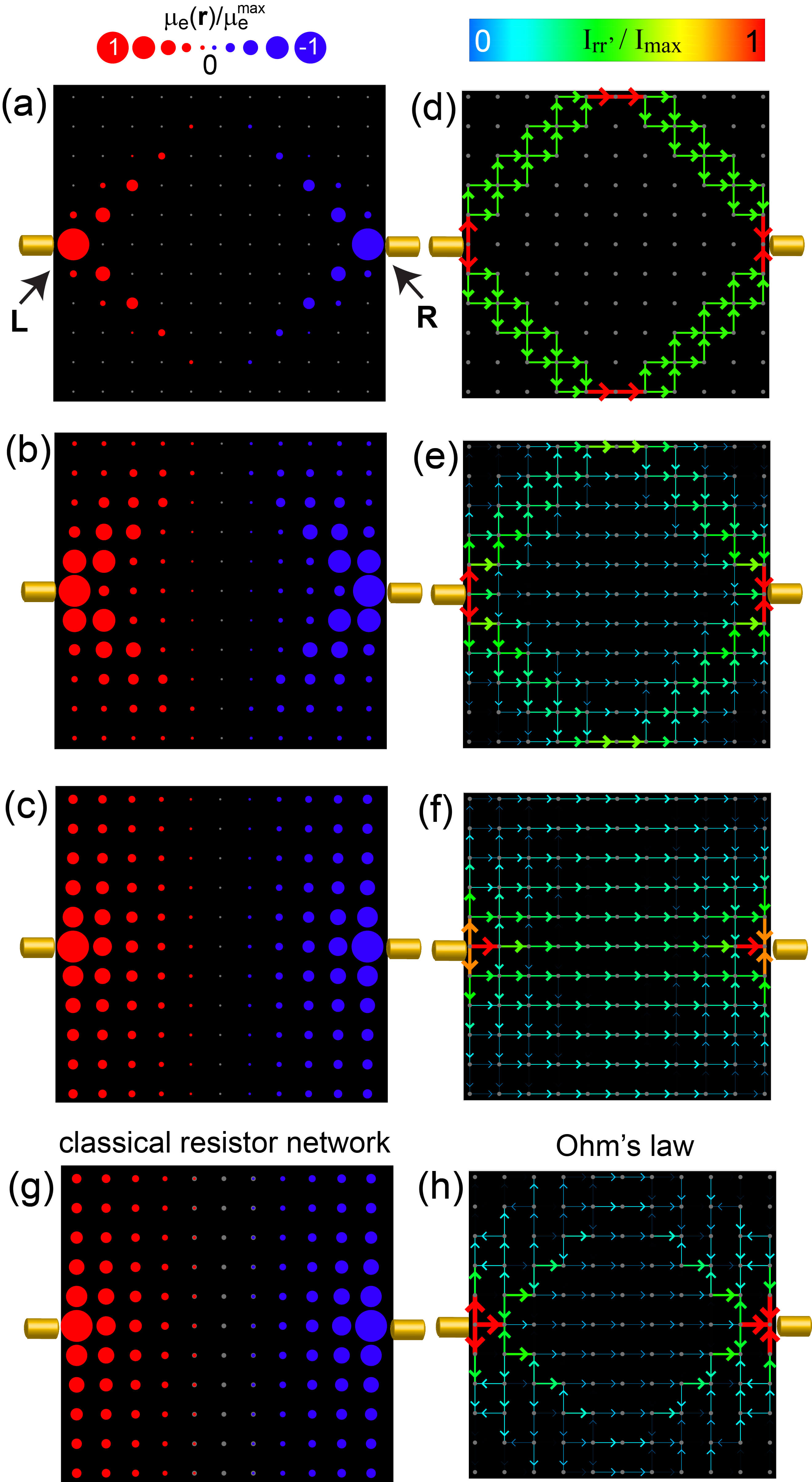}
\caption{Network with $N_x=N_y=11$: normalized $\mu_e({\bf r})/\mu_e^{max}$ for (a) $\zeta=0.01t^2$, (b) $\zeta=0.5t^2$, and (c) $\zeta=500 t^2$, and (c)-(e) corresponding normalized current pattern $I_{\bf r, r^\prime}/I_{max}$ for $T=0$, $t_c=t$ and $\mu_{L,R}=\pm 0.05t$. (g) $\mu_e({\bf r})$ in a classical resistor network connected to two narrow leads. (h) $I_{\bf r, r^\prime}$ obtained from (b) using Ohm's law with constant $\sigma_0$. $\mu_e({\bf r})$ at sites ${\bf L,R}$ in (a) and (b) has been divided by a factor 15 and 4, respectively, for clarity. }
\label{fig:spatial_V}
\end{figure}

When different chemical potentials, $\mu_{L,R}$ are applied to the left and right leads, a non-zero current flows through the network. The resulting spatial current pattern, $I_{\bf r,r^\prime}$ inside the network can be computed using the non-equilibrium Keldysh Green's function formalism \cite{Kel65,Ram86,Car71,Morr16}. At the same time, the current between the STP tip and a site ${\bf r}$ in the network in the weak tunneling limit is given by \cite{Bev14}
\begin{align}
I_{tip}({\bf r}) &= - 2 \frac{g_s e}{\hbar} N_0 t_{tip}^2 \int_{-\infty}^\infty
\frac{d \omega}{2 \pi} \left\{ \frac{{\rm Im}G^<({\bf r,r},\omega)}{2} \right. \nonumber \\
& \left. + n_F^{tip}\left[\omega - e V({\bf r}) \right] {\rm Im}G^r({\bf r,r},\omega)  \right\}
\label{eq:Itip}
\end{align}
where $G^{<,r}({\bf r,r},\omega)$ are the full local lesser and retarded Green's functions, $n_F^{tip}$ is the Fermi distribution function of the tip, and $V(\bf r)$ is the potential in the tip with respect to the network [for a detailed discussion of $G^<({\bf r,r},\omega)$, see Ref.~\cite{Morr16}]. To obtain the electro-chemical potential, $\mu_e({\bf r}) = e V(\bf r)$ via STP, $V(\bf r)$ is adjusted at every site ${\bf r}$ such that $I_{tip}({\bf r})=0$.

In Fig.~\ref{fig:mu_cuts}(c), we present the evolution of $\mu_e({\bf r})$ along the center row of the network in Fig.~\ref{fig:mu_cuts}(b) with increasing $\zeta$. In the non-interacting quantum limit, $\zeta=0$,  the  chemical potential abruptly changes at the lead-network interface, and is constant inside the network. This interface resistance limits the network's conductance to the   quantum of conductance \cite{Morr16}. With increasing $\zeta$, the resulting electronic dephasing leads not only to a varying $\mu_e({\bf r})$ inside the network, but also to an evolution in its spatial form, as shown in Figs.~\ref{fig:spatial_V}(a) - (c). To investigate the relation between $\mu_e({\bf r})$ and the corresponding spatial current pattern, $I_{\bf r,r^\prime}$ we plot the latter in Figs.~\ref{fig:spatial_V}(d) - (f) (for details of its calculation, see  Ref.\cite{Morr16}). For large $\zeta$, the spatial form of $\mu_e({\bf r})$ [Fig.~\ref{fig:spatial_V}(c) for $\zeta=500t^2$] and of $I_{\bf r,r^\prime}$ are that of a classical resistor network \cite{Wu04}, for which $\mu_e({\bf r})$ is shown in Fig.~\ref{fig:spatial_V}(g). In this case, $\mu_e({\bf r})$ and $I_{{\bf r,r^\prime}}$ (both obtained within the Keldysh formalism) are related by Ohm's law, $I_{{\bf r,r^\prime}}=\sigma({\bf r,r^\prime}) [\mu_e({\bf r}) - \mu_e({\bf r^\prime})]$, with the link conductivity between two neighboring sites being constant, i.e., $\sigma({\bf r,r^\prime}) = \sigma_0$. In the opposite limit of small $\zeta$, i.e., near the quantum limit, $\mu_e({\bf r})$ [Fig.~\ref{fig:spatial_V}(a)] shows a spatial form that is very similar to that of $I_{\bf r,r^\prime}$ [Fig.~\ref{fig:spatial_V}(d)], implying that $\mu_e({\bf r})$ can be used to spatially image regions of large current density. However, neither in this limit, nor in the crossover region between quantum and classical transport [as exemplified by $\zeta=0.5t^2$, Figs.~\ref{fig:spatial_V}(b) and (e)] are $\mu_e({\bf r})$ and $I_{\bf r,r^\prime}$ related by Ohm's law with a constant $\sigma_0$. To demonstrate this, we present in Fig.~\ref{fig:spatial_V}(h) a spatial plot of $I_{\bf r,r^\prime}$ obtained from $\mu_e({\bf r})$ in Fig.~\ref{fig:spatial_V}(b) [for intermediate $\zeta=0.5t^2$] using  Ohm's law with a constant $\sigma_0$. Not only does the resulting $I_{\bf r,r^\prime}$ not obey the continuity equation, but its spatial form is also qualitatively different from that of the actual current pattern shown in Fig.~\ref{fig:spatial_V}(e). We therefore conclude that the spatial current pattern $I_{\bf r,r^\prime}$ can only be extracted from $\mu_e({\bf r})$ via Ohm's law in the classical transport regime.

\begin{figure}
 \includegraphics[width=8cm]{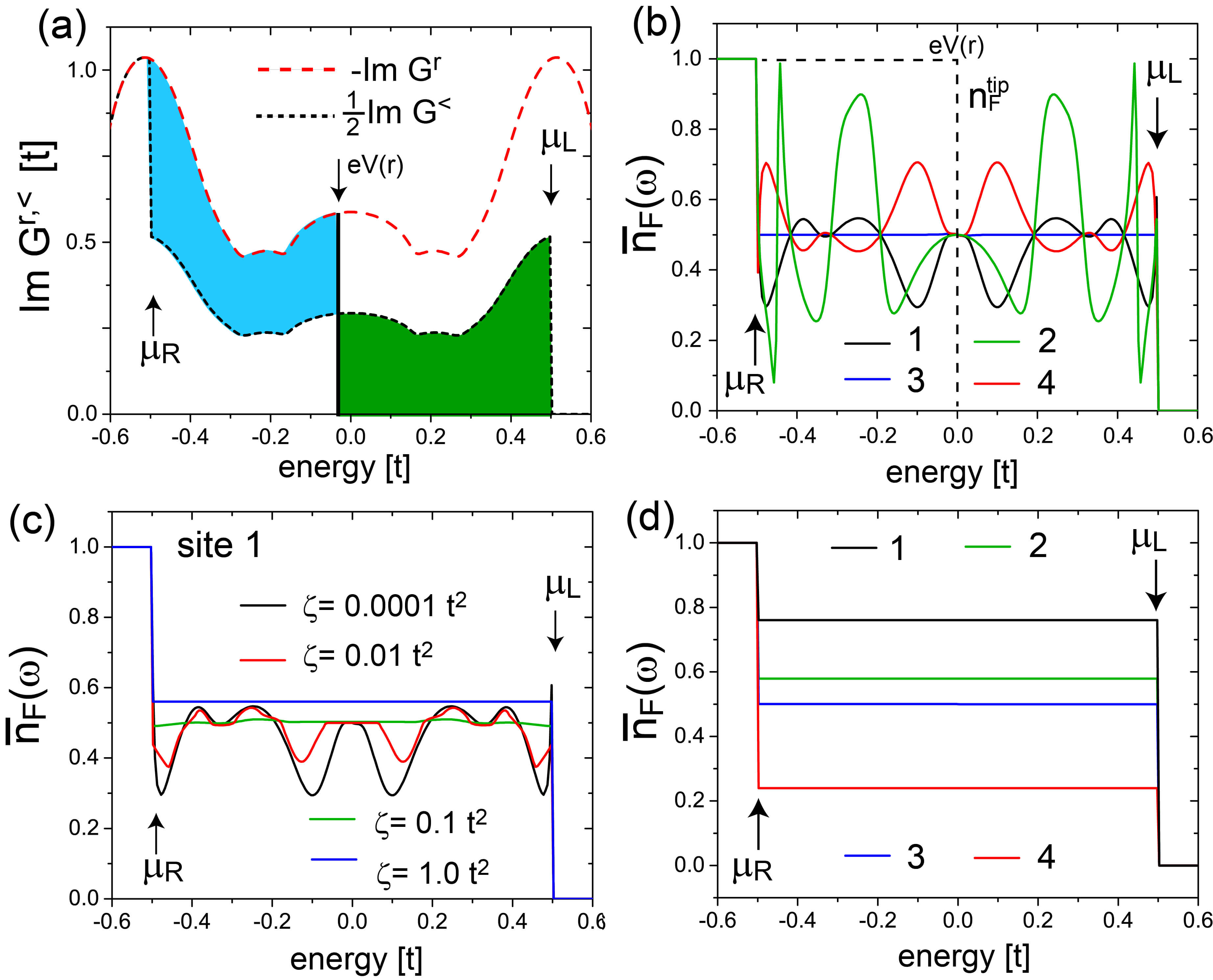}%
 \caption{(a) ${\rm Im}G^{r,<}$ at site 5 [see Fig.~\ref{fig:mu_cuts}(b)] for $\mu_{L,R} = \pm 0.5 t$ and $\zeta=0.1t^2$.  (b) ${\bar n}_F$ for $\mu_{L,R} = \pm 0.5 t$ and $\zeta=0.0001t^2$ at four different sites in the network [numbering corresponds to the sites in Fig.~\ref{fig:mu_cuts}(b)] and $n_F^{tip}$ (dashed line).(c) Evolution of ${\bar n}_F$ with increasing $\zeta$ at site 1. (d) ${\bar n}_F$ in the large $\zeta$ limit ($\zeta=100t^2$).}
 \label{fig:small_gamma}
 \end{figure}

Further insight into the nature of the local potential can be gained by considering a graphical solution of the condition $I_{tip}({\bf r})=0$ from Eq.(\ref{eq:Itip}). To this end, we present in Fig.~\ref{fig:small_gamma}(a) a plot of ${\rm Im} G^{<,r}$ for site 5 in Fig.~\ref{fig:mu_cuts}(b) and $\zeta = 0.1 t^2$. A closer analysis of Eq.(\ref{eq:Itip}) reveals that $V({\bf r})$ (for which $I_{tip}({\bf r})=0$) is determined by the condition that the area between $-{\rm Im} G^{r}$ and ${\rm Im} G^{<}/2$ for $\mu_R<\omega<eV({\bf r})$ (blue area) be equal to the area under ${\rm Im} G^{<}/2$ for $eV({\bf r})<\omega<\mu_L$ (green area). As previously pointed out \cite{Bev14}, these two areas can be interpreted as the currents flowing out of the tip into the right lead (blue area) and into the tip from the left lead (green area), respectively.
To gain insight into the physical processes involved, we define an effective out-of-equilibrium Fermi distribution function ${\bar n}_F$  in the network via $G^<({\bf r,r},\omega)=-2i{\bar n}_F(\omega) {\rm Im} G^r({\bf r},\omega)$. In equilibrium, ${\bar n}_F$ is the conventional Fermi distribution function. In Fig.~\ref{fig:small_gamma}(b) we present ${\bar n}_F$ for several sites in the network [the colors of the lines in Fig.~\ref{fig:small_gamma}(b) correspond to the colors of the circles in Fig.~\ref{fig:mu_cuts}(b)] for small $\zeta = 0.01 t^2$, together with the tip's Fermi distribution function, $n_F^{tip}$. As the network is out-of-equilibrium, ${\bar n}_F$ is modified from its equilibrium form in the energy range $\mu_{L}<\omega<\mu_{L}$ and varies greatly inside the network.  For $\mu_R<\omega<eV({\bf r})$, $n_F^{tip}=1>{\bar n}_F$, and these states carry a current from the tip into the network. On the other hand, for $eV({\bf r})<\omega<\mu_L$, one has $n_F^{tip}=0<{\bar n}_F$, and hence these states carry a current that flows from the network into the tip. For an appropriately chosen $V({\bf r})$, these two counterpropagating currents cancel, such that $I_{tip}({\bf r})=0$. We note that while ${\bar n}_F$ exhibits a strong energy dependence between $\mu_L$ and $\mu_R$ for small $\zeta$, this dependence becomes weaker with increasing $\zeta$, until ${\bar n}_F({\bf r},\omega) = {\bar n}^0_F({\bf r})$ is essentially constant for large $\zeta$. While the same qualitative evolution occurs at all sites in the network, the actual value of ${\bar n}^0_F$ in the limit $\zeta \rightarrow \infty$ depends on the location inside the network, as shown in Fig.~\ref{fig:small_gamma}(d) for the four sites indicated by filled circles in Fig.~\ref{fig:mu_cuts}(b). At the same time, ${\rm Im} G^r({\bf r},\omega)$ becomes nearly independent of energy for $\mu_R<\omega<\mu_L$, such that the graphic solution for finding $V({\bf r})$ discussed above now allows us to directly relate ${\bar n}^0_F$ and $\mu_e({\bf r})$ via
\begin{equation}
 \mu_e({\bf r})=\mu_R + {\bar n}^0_F({\bf r}) \left( \mu_L - \mu_R \right)
\end{equation}
The above discussion shows that the spatial dependence of $\mu_e({\bf r})$ is a truly non-equilibrium phenomenon, as it simply becomes equal to the network's uniform chemical potential in equilibrium where $\mu_{L,R}=0$. $\mu_e({\bf r})$ should also not be interpreted as representing a local equilibrium value, as the strong dependence of ${\bar n}_F$ on energy [see Figs.~\ref{fig:small_gamma}(b) and (c)] implies that ${\bar n}_F$ cannot be described by an equilibrium Fermi distribution function with a renormalized temperature or chemical potential.

\begin{figure}
 \includegraphics[width=8cm]{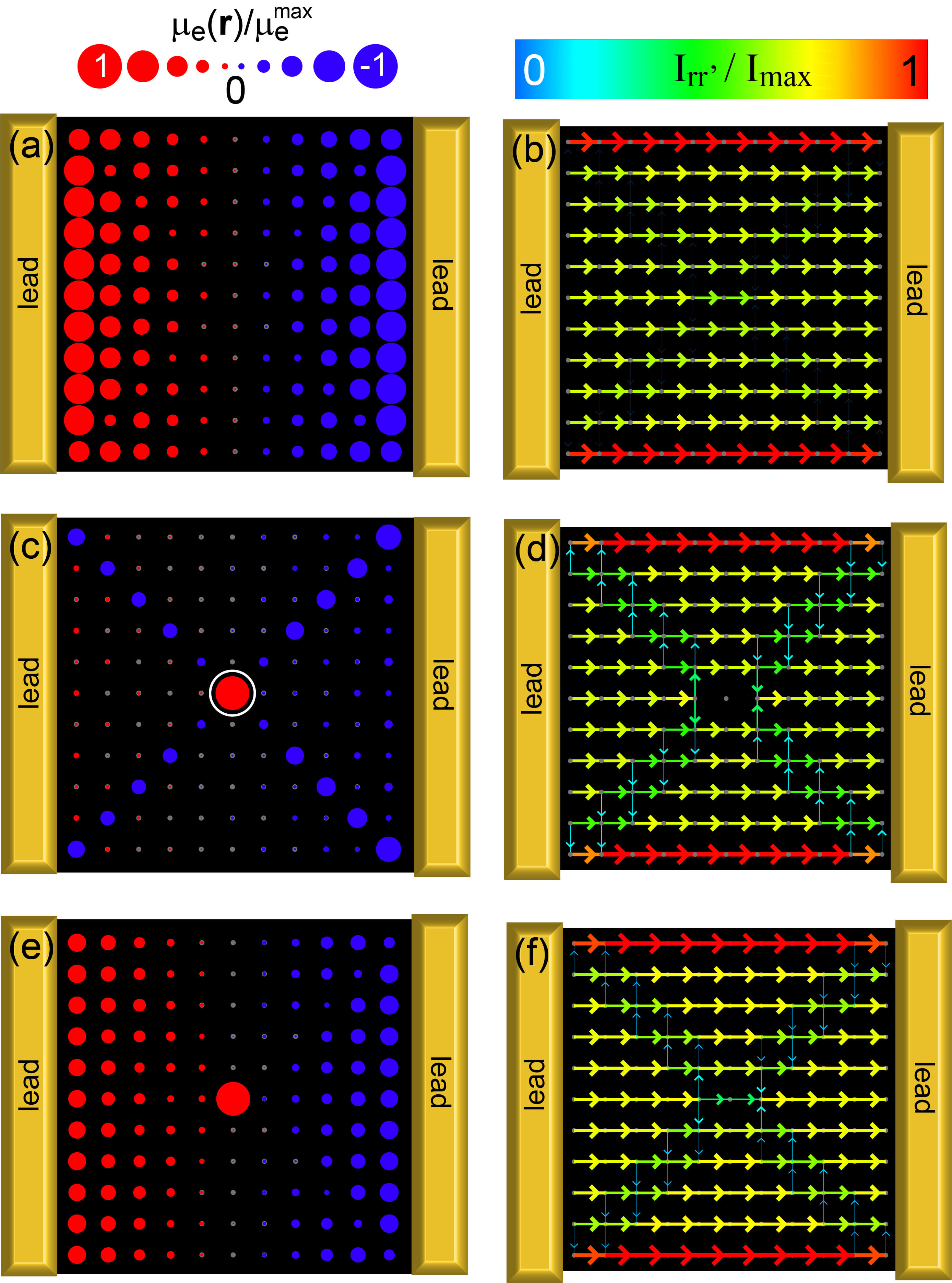}%
 \caption{Network connected to wide leads. (a) Normalized $\mu_e({\bf r})$ and (b) $I_{\bf r,r^\prime}$ for $\zeta = 0.01 t^2$. (c) - (f) Normalized $\mu_e({\bf r})$ and $I_{\bf r,r^\prime}$ for a network with a defect of $U_0=t$ located at the center [as indicated by an open white circle in (c)] and (c),(d) $\zeta = 0.01 t^2$, and  (e),(f)  $\zeta = 0.2 t^2$. $\mu_e({\bf r})$ at the defect site in (c) has been divided by a factor 3 for clarity. }
 \label{fig:widelead}
 \end{figure}

We next investigate the behavior of $\mu_e$ around defects, and to this end consider a network connected to wide leads [see Fig.~\ref{fig:widelead}]. In Figs.~\ref{fig:widelead}(a) and (b) we present the spatial form of $\mu_e({\bf r})$ and corresponding $I_{\bf r,r^\prime}$ near the ballistic quantum limit for a wide-lead network without a defect.
The current shows a very weak variation in magnitude inside the network, with the largest changes occurring along the edges, while the potential exhibits a variation across the network that is much more uniform than in the narrow lead case [see Fig.~\ref{fig:spatial_V}(a)]. The addition of a non-magnetic defect in the center of the network leads to significant changes in $\mu_e({\bf r})$ and $I_{\bf r,r^\prime}$ [see Figs.~\ref{fig:widelead}(c) and (d)] that extend throughout the entire network, and are predominantly confined to the lattice diagonal. This is a direct consequence of the Fermi surface's large degree of nesting [see Fig.~\ref{fig:mu_cuts}(c)] and a Fermi velocity along the diagonal direction in the Brillouin zone.  With increasing $\zeta$ , the effects induced by the defect in $\mu_e({\bf r})$ and $I_{\bf r,r^\prime}$ are reduced in amplitude [see Figs.~\ref{fig:widelead}(e) and (f)], and become spatially more confined to the immediate vicinity of the defect, indicating the crossover from non-local transport in the quantum limit, to local transport in the classical limit \cite{Morr16}.

\begin{figure}
 \includegraphics[width=8cm]{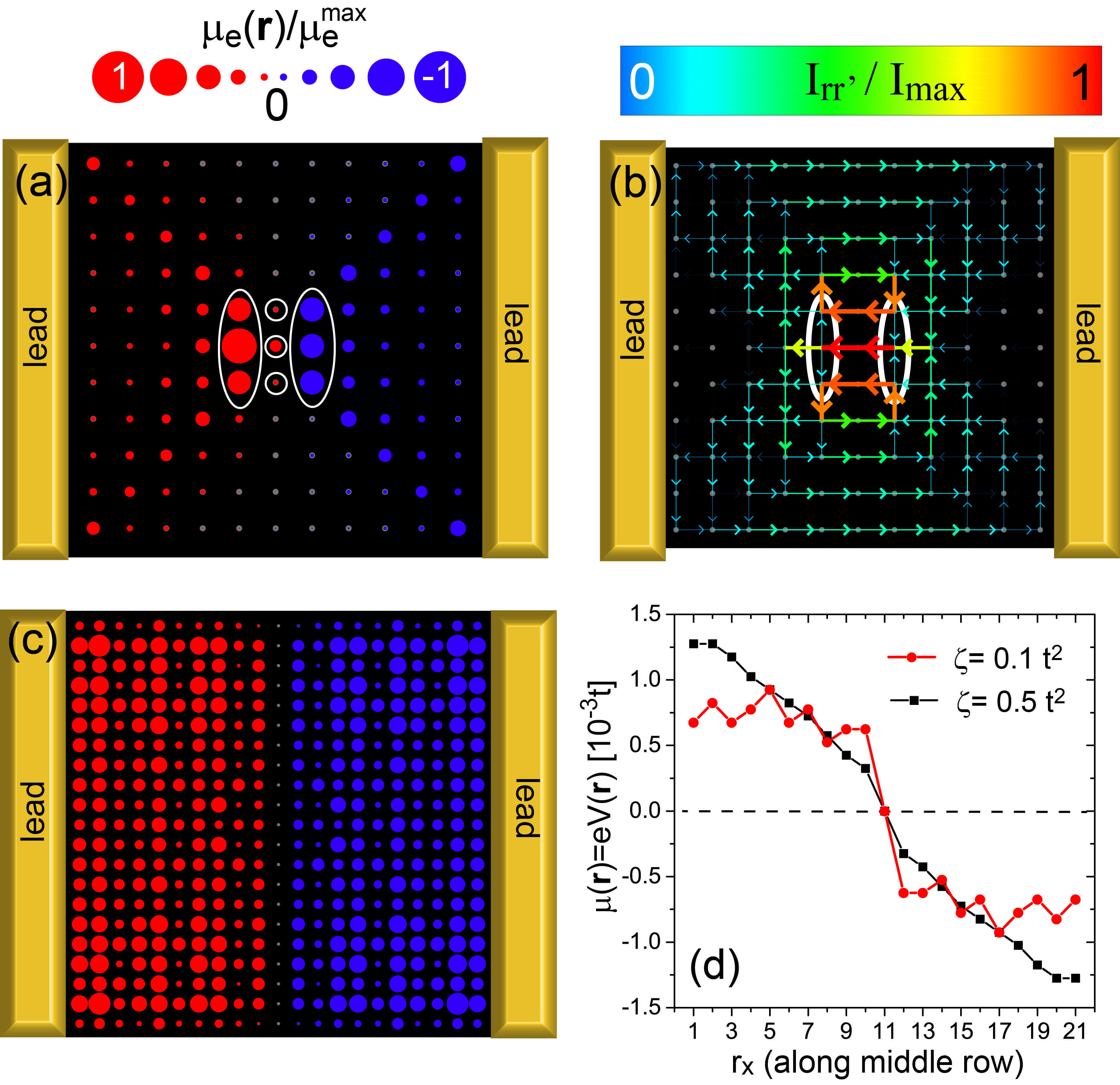}%
 \caption{Network connected to wide leads with three defects [as indicated by open white circles in (a)] of scattering strength $U_0=3t$.  (a) Normalized $\Delta \mu_e({\bf r})$ and (b) $\Delta I_{\bf r,r^\prime}$ for $\zeta = 0.5 t^2$. (c) Normalized $\mu_e({\bf r})$ for a network with $N_x=N_y=21$ and different chemical potentials in the left ($\mu=+t$) and right ($\mu=-t$) parts of the network, $\mu_{L,R}=\pm 0.01t$ and $\zeta=0.1t^2$. (d) Line cut of $\mu_e({\bf r})$ along the center row of (c).}
 \label{fig:RRD}
 \end{figure}

To visualize the formation of a residual resistivity dipole \cite{Lan57,Lan75}, we present in Figs.~\ref{fig:RRD}(a) and (b) the changes induced in the electro-chemical potential, $\Delta \mu_e({\bf r})$, and in the spatial current pattern, $\Delta I_{\bf r,r^\prime}$, respectively, by placing three defects [see small white circles in Figs.~\ref{fig:RRD}(a)] in the center of the network.
The spatial form of $\Delta \mu_e({\bf r})$ reveals the dipole nature of the induced changes, with an enhancement (suppression) of $\mu_e({\bf r})$ towards the lead with the higher (lower) chemical potential, thus demonstrating the existence of a defect-induced residual resistivity dipole. Interestingly enough, the spatial form of $\Delta I_{\bf r,r^\prime}$ [see Fig.~\ref{fig:RRD}(b)] is that of field lines associated with the presence of a dipole. This becomes particulary evident when we indicate the regions with the largest $\Delta \mu_e({\bf r})$ (see white ellipses next to the defects) in the plot of $\Delta I_{\bf r,r^\prime}$. We therefore conclude that the relation between the defect-induced changes in $\mu_e({\bf r})$ and $I_{\bf r,r^\prime}$ is that of dipole charges and their associated field lines. Finally, to explore the form of $\mu_e({\bf r})$ near interfaces or step edges, we apply different chemical potentials to the left ($\mu=+t$) and right ($\mu=-t$) parts of a network. The resulting $\mu_e({\bf r})$ shown in Fig.~\ref{fig:RRD}(c) and (d), exhibit not only as expected a sharp drop at the center of the network where the change in chemical potential occurs, but also spatial oscillations that extend all the way back to the leads. This is reminiscent of the spatial oscillations found near step edges in \cite{Wil15}. With increasing $\zeta$, this sharp drop is smoothed out, leading to a mare gradual variations of $\mu_e({\bf r})$ across the network [Fig.~\ref{fig:RRD}(d)].

In summary, we identified the spatial relation between the electrochemical potential and the current patterns over the entire range from quantum to classical transport. These two quantities show similar spatial patterns near the quantum limit, but are related by Ohm's law only in the classical regime. We showed that defects induce  a Landauer residual resistivity dipole in $\mu_e({\bf r})$,  with the spatial form of the concomitant $\Delta I_{\bf r,r^\prime}$ representing the field lines associated with the dipole. It would be interesting to use a similar approach to investigate the relation between heat currents and local temperature measurements out-of-equilibrium \cite{Ber15,Sha16}.

\begin{acknowledgments}
We would like to thank M. Beasley, M. Graf and M. Wenderoth for stimulating discussions.  This work was supported by the U. S. Department of Energy, Office of Science, Basic Energy Sciences, under Award No. DE-FG02-05ER46225.
\end{acknowledgments}

\end{document}